# *Fermi Liquid Model for Hadrons in Medium*

**A. Bhattacharya [1*], S. Pal [2], R. Ghosh [3], B.Chakrabarti [4]**

[1*,3] Department of Physics, Jadavpur University,
Kolkata 700032, India.
[2] Department of Physics, Basanti Devi College, Kolkata, India.
[4] Department of Physics, Jogamaya Devi College, Kolkata, India.
1* corresponding author

**Abstract:** A Fermi liquid model for hadrons has been suggested for the hadrons in medium. The hadrons are supposed to behave like quasi particle as Fermi excitation while in the medium and the effective mass of the hadrons have been estimated using Fermi liquid model. Considering a momentum dependent potential inside the medium to describe the interaction, the effective masses of the hadrons are estimated. The temperature dependence of effective masses has also been studied. The possibility of describing masses of the quarks as Fermi excitation has been investigated. Compressibility, specific heats, density of states in medium has been studied. The potential depth for light and singly heavy baryons in medium has been extracted. The results are found to be very interesting and compared with the other studies available in literature.

PACS number(s): 12.39.-x, 12.39.Pn, 12.40.-y, 12.90.+b

## *Introduction:*

The Landau Fermi liquid theory is an phenomenological approach to strongly interacting normal fermi system at small excitation energies. It is a model which suggests a point to point correspondence between low energy excitation of non interacting Fermi gas [1]. The model has been widely used to study the properties of liquid He-3, electron in metal and nuclear matter. It gives an effective description of low energy elementary excitations like the quasiparticles in crystal lattice. The model is found to be successful in describing some aspects of QCD, quark and hadronic matter. Heugten et al [2] have used fermi liquid theory to study the specific heat, quark effective mass and thermodynamic potential of QCD for imbalance quark matter. Schafer [3] has made a brief review of Fermi liquid in the context of dilute neutron matter, cold atomic gases and dense quark matter. He has discussed the Fermi parameter $F_0$, $F_1$ of neutron matter with fermi momentum $K_F$ and have observed that in the range 0.6 fm$^{-1}$ < $P_F$ < 1.4 fm$^{-1}$ the effective mass exceeds unity whereas Nascibene et al [4] have observed $\frac{m^*}{m} = 1.3 \pm 0.03$. Friman et al [5] have investigated the properties of hadronic matter derived from a chiral lagrangian with Brown-Rho (BR) scaling whereas Song et al [6] have applied (BR) scaling to describe fluctuation under extreme condition relevant to neutron star. The possibility of spontaneous magnetization and magnetic properties of quark matter evaluating magnetic susceptibility within Fermi liquid have been discussed by Tutsumi et al [7]. Kitzawa et al [8] have investigated the quasiparticle picture of quarks with temperature and have shown that a non Fermi liquid behavior of matter develops near the critical point. Chen [9] has investigated Landau Fermi liquid parameters by non linear scaling transformation. They have obtained Landau parameters $F_0$ and $F_1$ for strongly interacting Fermion system. They have suggested $\frac{m^*}{m} = \frac{10}{9}$ as universal constant at unitarity.

In the present work we have investigated the properties of hadrons in nuclear matter with Fermi liquid model type excitation of the particles in medium. We follow the qualitative picture developed to describe a fermion in medium. We have estimated the effective mass of the hadrons in medium considering the fact that the hadrons behave like fermi excitation while in medium and have estimated the dressed mass considering a suitable potential. The ratios of masses, specific heats, compressibility, binding energies etc. have been studied.

## *Formulation:*

In Fermi liquid theory [1] the real fermi gas are not point particles and if one moves other must give the way. If a weak potential is assumed, the excitation spectrum remains same qualitatively similar as in free fermi gas with a shift of energy [10] so free fermi sphere is associated with another particle at momentum p with p > $p_F$ and forms excited state of ideal gas with modified mass $m^*$ which has been defined as effective mass. The modified density of states can be represented as:

$$g(\epsilon_F) = \frac{m^* p_F}{\pi^2 \hbar^3} \tag{1}$$

The specific heat also differs from ideal value. So the low energy properties of fermi liquid can be understood as an ideal gas with difference that the effective mass appears in stead of particle mass m. The energy can also be expressed as: [11]

$$\epsilon(p \simeq p_F) = \frac{p^2}{2m} + V(p) = \frac{p^2}{2m} + const. \tag{2}$$

The assumption of this expression (2) is that for quasi particle with momentum p ≃ $p_F$, the modification of V (p) brought into the quantity $\in$ (p) by the presence of inter particle interaction in the liquid and may be represented by a constant term while the kinetic energy $\frac{p^2}{2m}$ is modified so as to replace the particle mass by $m^*$, the effective mass.
Differentiating expression (2) with respect to p and setting p = $p_F$ we obtain [11]

$$\frac{1}{m^*} = \frac{1}{m} + \frac{1}{p_F}[\frac{dV(p)}{dp}]_{p=p_F} \qquad (3)$$

The quantity $m^*$ in particular determines the low temperature specific heat of the Fermi liquid. The ratios of excited fermi gas to that of an ideal fermi gas is precisely equal to the ratio $\frac{m^*}{m}$ so that [11];

$$\frac{C_{V_{real}}}{C_{V_{ideal}}} = \frac{m^*}{m} \qquad (4)$$

At low temperature only momentum close to $p_F$ are important and in the vicinity of the fermi surface the particle mass is replaced by effective mass $m^*$. The effective mass of a quasi particle with momentum $p_F$ can be expressed as:[11]

$$m^* = \frac{p_F}{U_F} = \frac{p_F}{(\frac{\delta\epsilon}{\delta p})_{p=p_F}} \qquad (5)$$

or

$$\frac{1}{m^*} = \frac{1}{m} + \frac{1}{p_F}[\frac{dV(p)}{dp}]_{p=p_F} \qquad (6)$$

where V(p) is the potential describing the interaction.
We have used similar type of model for hadrons in medium. We assume that the mass of the hadrons gets modified as it propagates through the medium due to the interaction. To describe the interaction we consider a momentum dependent potential as [12];

$$V(r, p^2) = V_0 e^{-\gamma(\frac{p^2}{m})} v(r) \qquad (7)$$

Considering potential as a function of momentum only $v(r)$ constant, the above expression can be recast as:

$$V(p^2) = V_0 e^{-\gamma\left(\frac{p^2}{m}\right)} \tag{8}$$

Where $\gamma \simeq \frac{1}{V_0}$ Combining equation (6) and (7), we come across an expression for $m^*$ such as:

$$\frac{1}{m^*} = \frac{1}{m} - \frac{2}{m} e^{-\gamma\left(\frac{p^2}{m}\right)} \Big|\ p = p_F \tag{9}$$

We can estimate the effective mass of the hadron with the knowledge of fermi momentum. To estimate the fermi momentum we recall one of our previous works [13]. The Fermi distribution formula runs as:

$$n(r) = \frac{N}{V} = \frac{1}{\pi^2 \hbar^3} \int_0^\alpha \frac{p^2 dp}{1 + e^{(\epsilon - \mu)/KT}} \tag{10}$$

with chemical potential $\mu = ar$ and $\epsilon = \frac{p^2}{2m}$, we have obtained an expression relating the radius parameter of the corresponding hadrons and fermi momentum [13]. At T=0 we have;

$$p_F^2 = \frac{5}{3} a.m.r_0 \tag{11}$$

where $r_0$ is the radius parameter of the particle, m is the constituent mass of the corresponding hadron. The fermi momentum of the particle have been estimated with a = 0.02 GeV$^2$ [14] and with the input of radius parameter for different hadrons obtained in the literature [15-21]. The radius parameters for different - hadrons, respective fermi momentum are furnished in the table-

I, II, III, IV for light, heavy, doubly heavy and triply heavy baryons along with the ratios of modified mass and original mass of the hadrons as it propagates through the medium with interaction representing in (7) with $\gamma = \frac{1}{V_0} = 25$ MeV representing approximate potential - depth. Variation of $\frac{m^*}{m}$ with temperature have been studied for light and single heavy baryons and displayed in Fig-I and Fig-II considering $T_C$ = 170 GeV. We have also estimated the $\gamma$ for light and singly heavy baryons from the study of variation of $\frac{m^*}{m}$ with $p_F^2$ and displayed in the Fig-III to XIV. The values of $\gamma$ for different hadrons have been extracted from the linear part of the graphs which gives us an approximate idea about the potential well of the corresponding hadrons. The effective masses for nucleons have estimated and displayed in Table-V along with the compressibility of nucleons.

The expression for compressibility of nucleons runs as:

$$\kappa = \frac{1}{3} r_0^2 \frac{\delta^2 E}{\delta r^2} \tag{12}$$

We have estimated compressibility of proton as $\kappa_p$ = 0.314 GeV and that of neutron as $\kappa_n$ = 0.258 GeV and 0.381 GeV for two different radius using Fermi liquid excitation model for nucleon when nucleon is in medium.

The Landau parameter $F_1^S$ is related to the effective mass as:

$$\frac{m^*}{m} = 1 + \frac{F_1}{3} \tag{13}$$

We have estimated the values of Landau parameter $F_1^S$ for different hadrons and displayed in Table-VI.

## *III Discussions and Conclusions:*

In the present work we have studied the properties of hadrons in medium in an analogy with the Fermi liquid model where the effect of the medium is incorporated via effective mass approximation. We have assumed that a hadron in medium behaves like a Fermi excitation incorporating the many body interaction in the system and behaves like a quasi particle with an effective mass different from constituent mass. A momentum dependent potential has been assumed to simulate many body interaction in the medium. The results are displayed in the tables. It has been observed that the ratios of effective mass and constituent masses are more than one and vary from 1.1 to 1.3. We also have observed larger values for some of the hadrons. In light sector (Table-I) for $\lambda^0$ ratio is ~ 1.6. We have also observed larger value of the ratios for hadrons like $\lambda_b^0, \Sigma_b^0, \Xi_b^0, \Xi_{cc}^+, \Omega_{cc}^+, \Omega_{cb}^+, \Omega_{ccc}, \Omega_{bbb}$. It is evident from (4) that larger value of effective mass corresponds to larger value of specific heat which can be attributed to the enhanced value of the effective mass of quasi particles in the medium. It has been pointed out by Guenault [22] that for liquid $^3$He $\frac{m^*}{m}$ = 2.8. The value of Landau parameters have been shown in Table-VI and the values are found to be varying from 0.051 from 11.04. In non interacting assembly of gases the values of all F are zero whereas the larger values of $F_1$ indicate strongly interacting system. It may be pointed out that for liquid He $F_1$ is found to be 5.4 at zero pressure and found to be as large as 14 at very high pressure [22]. It is interesting to observe here that the effective mass exceed unity in all cases where the Fermi momentums are found to be in the range ~ 0.3GeV to 1.3GeV for the hadrons. It may be pointed out that Schafer [3] has observed that the effective mass exceeds unity while the Fermi momentum is in the range $0.6 fm^{-1} < P_F < 1.4 fm^{-1}$ while investigating the properties of neutron matter, quark matter and cold atomic gas in the context of fermi liquid theory. Nascimbene et al [4] have shown that the low temperature thermodynamics of strongly interacting normal phases are well described by Fermi liquid theory and the normal phase behaves as an mixture of two ideal gas although they have strong interaction which can be represented by a bare majority atoms and anon interacting gas of dressed quasi particles. They have obtained $m^*/m = 1.13 \pm 0.03$ for two component fermi gas.

In fig I and II we have observed that the there is a linear zone for all the $\frac{m^*}{m}$ vs $\frac{T}{T_C}$ graphs for all the light baryons and the range is found to be ~ 0.34 to 0.69 $\frac{T}{T_C}$. We have obtained a linear equation like $\frac{m^*}{m} = -0.126 \frac{T}{T_C}$ for light baryons except $\Xi^-$ particle where the slope is found to be ~ -0.069. The equation for singly heavy barons are found to be $\frac{m^*}{m} = -2.18 \frac{T}{T_C}$ in the linear zone from Fig-II where the linear zone runs from 0.4 to 0.65 $\frac{T}{T_C}$. So it may be suggested that in the aforesaid zone the equation have an universal applications and can be applied for all light baryons ( Fig-I) and heavy baryons( Fig-II) respectively. Fig-III to XII shows our study of effective mass vs fermi momentum. From equation (9), we approximately derive the relation which runs as $m\left(1 + \frac{m}{m^*}\right) = 2\gamma p_F^2$. We have plotted the graphs for each of light and heavy baryons and extracted the values of $\gamma$ for different light and heavy baryons which gives us an idea about the approximate available well depth for the particles. For $\Lambda$ and $\sigma$ the value is found to be ~ 25 GeV$^{-1}$ whereas for $\Xi$ and $\Omega$ the value is found to be  ~ 35 GeV$^{-1}$ which yield potential well depth $V_0$ i,e $\frac{1}{\gamma}$ as ~ 40 Mev and 28.5 Mev respectively. For singly heavy baryons the value of $\gamma$ has been obtained as ~ 42GeV$^{-1}$ which yields $V_0$ ~ 23.8 MeV. Kohono et al [23] have studied the momentum dependence of potential well depth of $\lambda$ using SU (6) quark model, they have estimated the single particle potential in nuclear matter and have obtained potential - depth > 40 MeV. The experimental value of potential well depth in nuclear matter for $\lambda$ is suggested to be -30 MeV [24]. Dover et al [25] have observed $\Xi$ well depth ~ 21-24 MeV from the emulsion date whereas relativistic potential predicts the value as -28 MeV [26]. Zhao et al [27] has studied potential well depth of $\Sigma$ in nuclear matter on surface gravitational red shift of proto neutron star and have observed that as the potential well depth increases from - 35 to + 35 MeV surface gravitational red shift increases and contribution is close to one percent. In the current investigation we have extracted the potential well depth for $\lambda$ as 40 MeV which closely agree with the value of Kohono et al [23]. The potential well depth for $\Xi$ is found to be 28.5 MeV in the current work which is in agreement with the relativistic potential calculation obtained by Bielich et al [26]. For heavy baryons like $\lambda_c^+$, $\lambda_b$, $\Xi_c^+$, we have obtained the value as ~ 24MeV.

In the present work we have studied the modification of properties of baryons in the medium in the context of fermi liquid model. Enhancement in masses have been estimated which gives us a number of interesting facts regarding the properties of the baryons in the medium. We have extracted the values of available well depths which in fact gives us idea about the binding energy of the particles in medium. It is interesting to note that the hadrons in the medium behaves like a strongly interacting system and can be represented as a Fermi excitation.

*Table I*: $m^*/m$ for Light sector Hadrons.

| Hadron | Radius in GeV$^{-1}$ | Fermi momentum ($P_F$) in GeV | $m^*$ in GeV | $\dfrac{m^*}{m}$ |
|---|---|---|---|---|
| $\lambda^0$(uds) | 1.9895 | 0.2889 | 2.037 | 1.616 |
| $\Sigma^-$(dds) | 3.65 | 0.3915 | 1.393 | 1.1055 |
| $\Xi^-$(dss) | 3.3 | 0.3979 | 1.6512 | 1.1466 |
| $\Omega^-$(sss) | 2.9 | 0.3957 | 1.972 | 1.217 |

**Table II**: $m^*/m$ for Single Heavy Baryons.

| Hadron | Radius in GeV$^{-1}$ | Fermi momentum ($P_F$) in GeV | $m^*$ in GeV | $\dfrac{m^*}{m}$ |
|---|---|---|---|---|
| $\lambda_c^+$(udc) | 5.727 | 0.6582 | 2.3094 | 1.017 |
| $\lambda_b^0$(udb) | 1.481 | 0.4928 | 11.775 | 2.3933 |
| $\Sigma_c^+$(udc) | 3.386 | 0.5061 | 2.5766 | 1.135 |
| $\Sigma_b^0$(udb) | 1.253 | 0.4533 | 16.6389 | 3.382 |
| $\Xi_c^0$(dsc) | 2.404 | 0.4430 | 3.3557 | 1.3696 |
| $\Xi_b^0$(usb) | 1.12 | 0.4363 | 23.88 | 4.682 |
| $\Omega_c^0$(ssc) | 2.5 | 0.4681 | 3.5026 | 1.332 |
| $\Omega_b^-$(ssb) | 2.0124 | 0.5951 | 8.431 | 1.5967 |

**Table III**: $m^*/m$ for Doubly Heavy Baryons.

| Hadron | Radius in GeV$^{-1}$ | Fermi momentum ($P_F$) in GeV | $m^*$ in GeV | $\dfrac{m^*}{m}$ |
|---|---|---|---|---|
| $\Xi_{cc}^{++}$(ucc) | 4.549 | 0.7243 | 3.6245 | 1.0475 |
| $\Xi_{cc}^+$(ccd) | 1.443 | 0.4079 | 8.673 | 2.5066 |
| $\Xi_{bb}^0$(ubb) | 3.917 | 1.0679 | 9.4876 | 1.0830 |
| $\Xi_{bb}^-$(bbd) | 3.039 | 0.9419 | 10.4155 | 1.189 |
| $\Xi_{cb}^+$(ucb) | 4.4017 | 0.9468 | 6.4391 | 1.0538 |
| $\Xi_{cb}^0$(cbd) | 1.9895 | 0.6365 | 9.8716 | 1.6156 |
| $\Omega_{cc}^+$(scc) | 1.3437 | 0.4037 | 10.486 | 2.88 |
| $\Omega_{cb}^0$(cbs) | 1.054 | 0.4701 | 37.214 | 5.9163 |
| $\Omega_{bb}^-$(bbs) | 2.4438 | 0.8533 | 12.097 | 1.3531 |

**Table IV**: $m^*/m$ for Triple Heavy Baryons.

| Hadron | Radius in GeV$^{-1}$ | Fermi momentum ($P_F$) in GeV | $m^*$ in GeV | $\dfrac{m^*}{m}$ |
|---|---|---|---|---|
| $\Omega_{ccc}$(ccc) | 1.5 | 0.4822 | 10.89 | 2.3419 |
| $\Omega_{bbb}$(bbb) | 1.25 | 0.7245 | 42.818 | 3.398 |
| $\Omega_{ccb}$(ubb) | 5 | 1.1029 | 7.5336 | 1.032 |
| $\Omega_{bbc}$(bbc) | 5 | 1.2877 | 10.2684 | 1.032 |

**Table V**: $m^*/m$ and compressibility for Nucleons.

| Nucleon | Radius in GeV$^{-1}$ | Fermi momentum ($P_F$) in GeV | $m^*$ in GeV | $\dfrac{m^*}{m}$ | Kappa(K) in GeV |
|---|---|---|---|---|---|
| Neutron | 4.7 | 0.4113 | 1.1247 | 1.041 | 0.258 |
| Neutron | 4 | 0.3795 | 1.1629 | 1.0768 | 0.381 |
| Proton | 4.375 | 0.3968 | 1.1394 | 1.055 | 0.314 |

**Table VI**: $m^*/m$, $F_1^{(s)}$, $g(\epsilon_F)$ and $v_F$.

| Nucleon | $\dfrac{m^*}{m}$ | $F_1^{(s)}$ | $g(\epsilon_F)$ in GeV$^2$ | $v_F$ |
|---|---|---|---|---|
| $\lambda^0$(uds) | 1.616 | 1.848 | 0.059 | 0.142 |
| $\Sigma^-$(dds) | 1.105 | 0.316 | 0.055 | 0.281 |
| $\Xi^-$(dss) | 1.147 | 0.439 | 0.067 | 0.241 |
| $\Omega^-$(sss) | 1.217 | 0.651 | 0.079 | 0.201 |
| $\lambda_c^+$(udc) | 1.017 | 0.051 | 0.154 | 0.285 |
| $\lambda_b^0$( udb) | 2.393 | 4.180 | 5.803 | 0.042 |
| $\Sigma_c^+$(udc) | 1.135 | 0.435 | 0.132 | 0.196 |
| $\Sigma_b^0$ (udb) | 3.382 | 7.146 | 0.764 | 0.027 |
| $\Xi_c^0$(dsc) | 1.369 | 1.109 | 0.151 | 0.132 |
| $\Xi_b^0$(usb) | 4.682 | 11.046 | 1.056 | 0.018 |
| $\Omega_c^0$(ssc) | 1.332 | 0.992 | 0.166 | 0.134 |
| $\Omega_b^-$(ssb) | 1.597 | 1.790 | 0.508 | 0.071 |

*Ackowledgement:*

Authors are thankful to University Grants Commission, New Delhi, India for financial assistance.

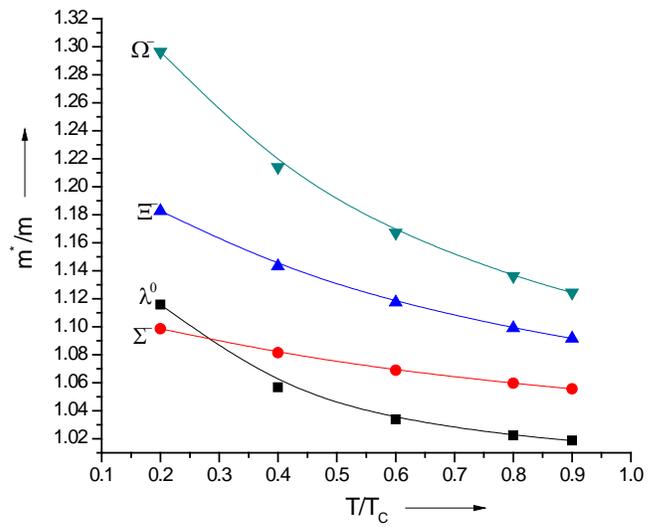

Figure 1. For Light Hadrons, $T/T_C$ Versus $m^*/m$

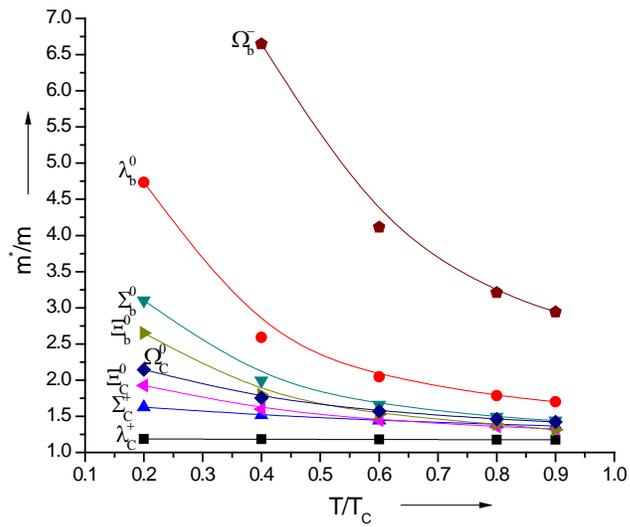

Figure 2. For Light Hadrons, $T/T_C$ Versus $m^*/m$

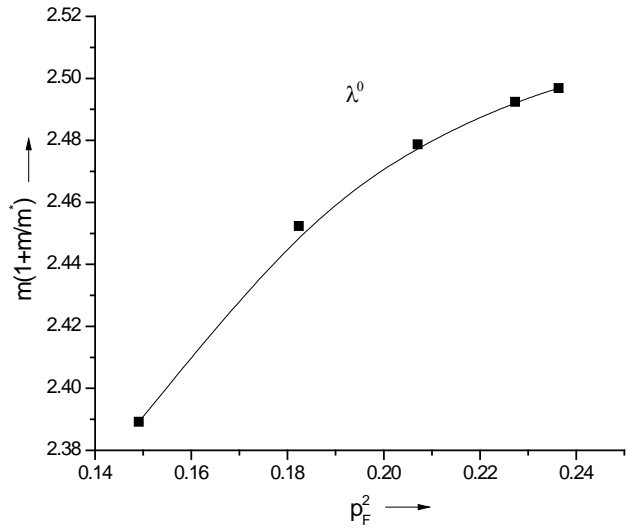

Figure 3. Variation of $p_F^2$ versus $m\left(1 + {m}/{m^*}\right)$

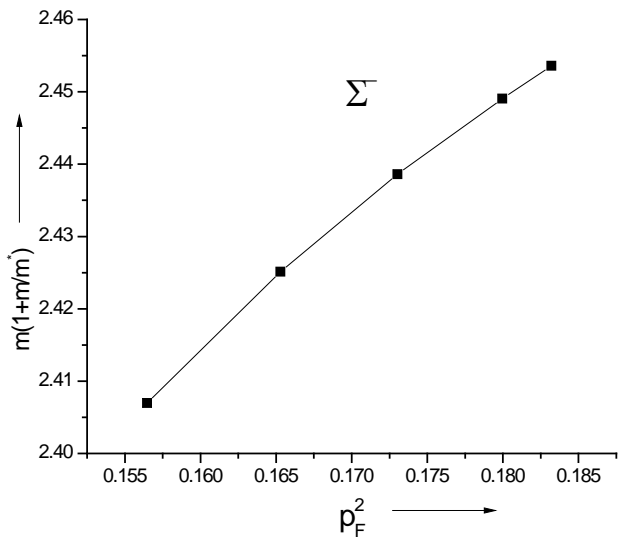

Figure 4. Variation of $p_F^2$ versus $m\left(1 + {m}/{m^*}\right)$

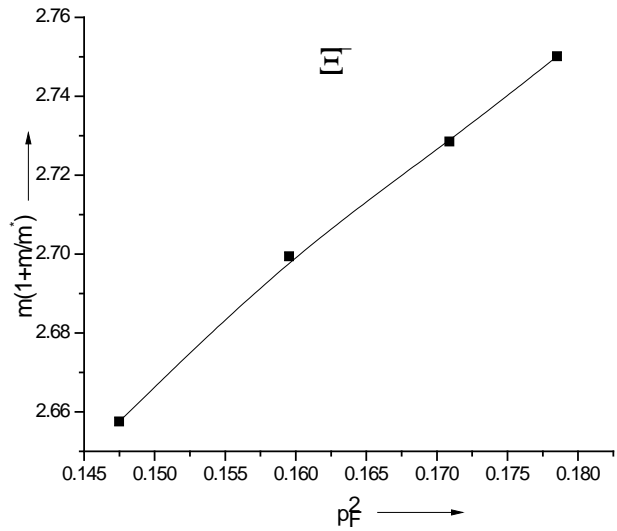

Figure 5. Variation of $p_F^2$ versus $m\left(1 + m/m^*\right)$

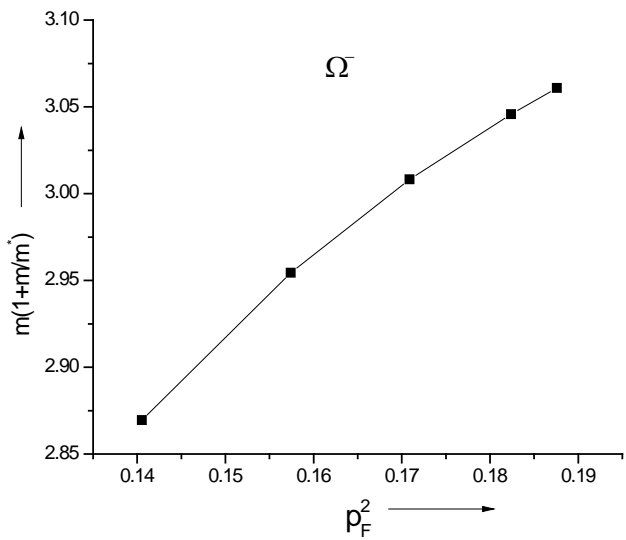

Figure 6. Variation of $p_F^2$ versus $m\left(1 + m/m^*\right)$

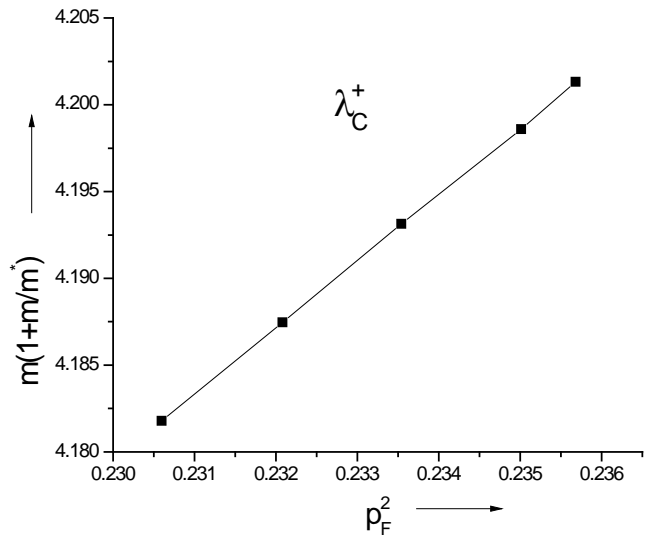

Figure 7. Variation of $p_F^2$ versus $m\left(1 + m/m^*\right)$

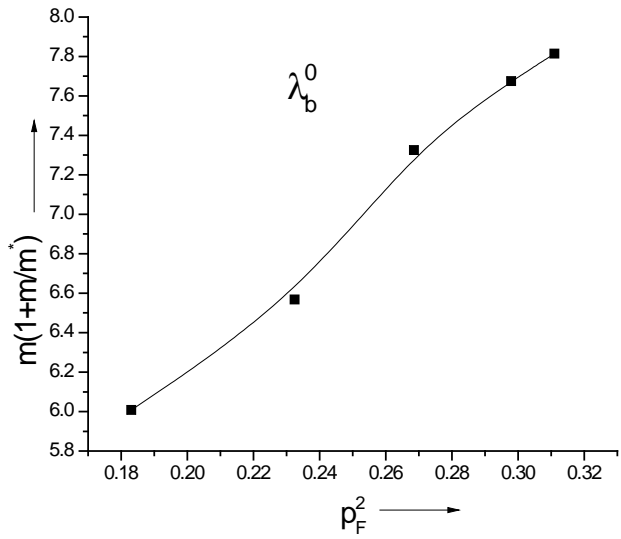

Figure 8. Variation of $p_F^2$ versus $m\left(1 + m/m^*\right)$

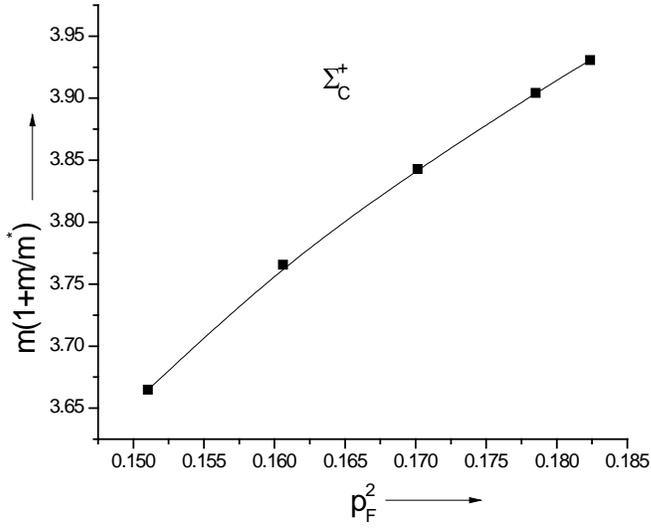

Figure 9. Variation of $p_F^2$ versus $m\left(1+ {m}/{m^*}\right)$

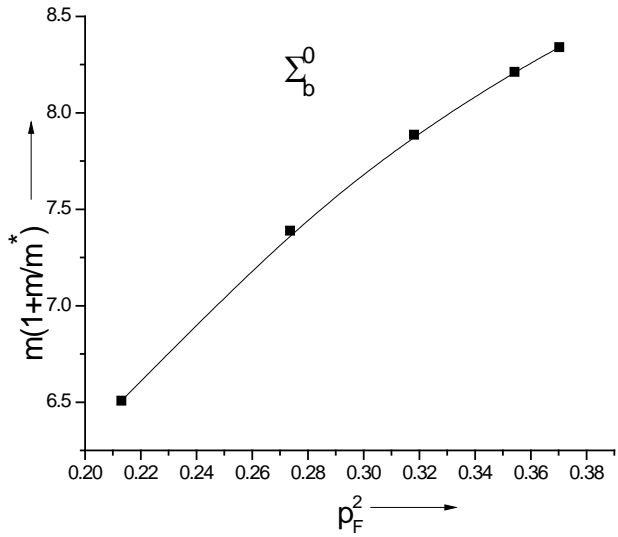

Figure 10. Variation of $p_F^2$ versus $m\left(1+ {m}/{m^*}\right)$

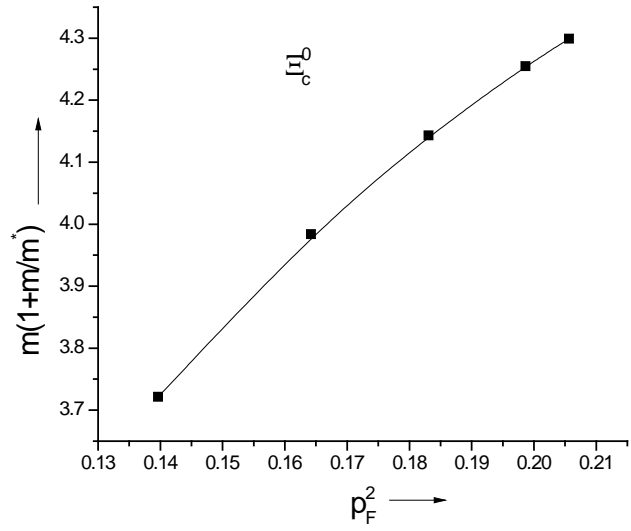

Figure 11. Variation of $p_F^2$ versus $m\left(1 + m/m^*\right)$

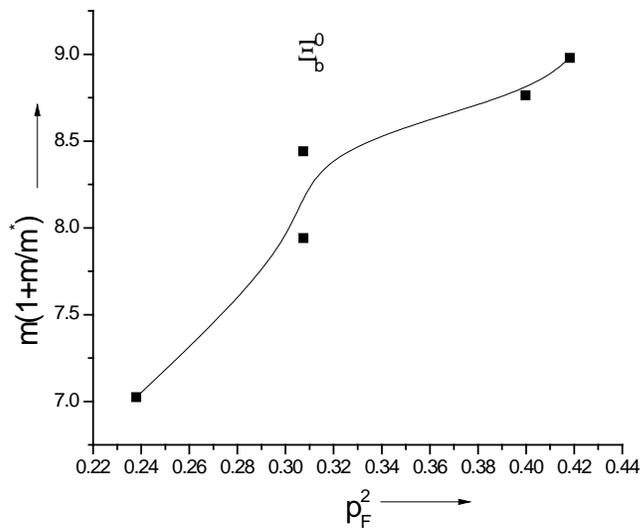

Figure 12. Variation of $p_F^2$ versus $m\left(1 + m/m^*\right)$

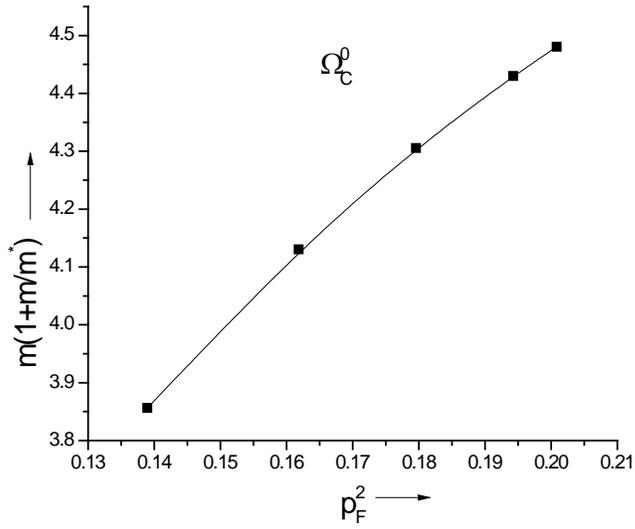

Figure 13. Variation of $p_F^2$ versus $m\left(1 + m/m^*\right)$

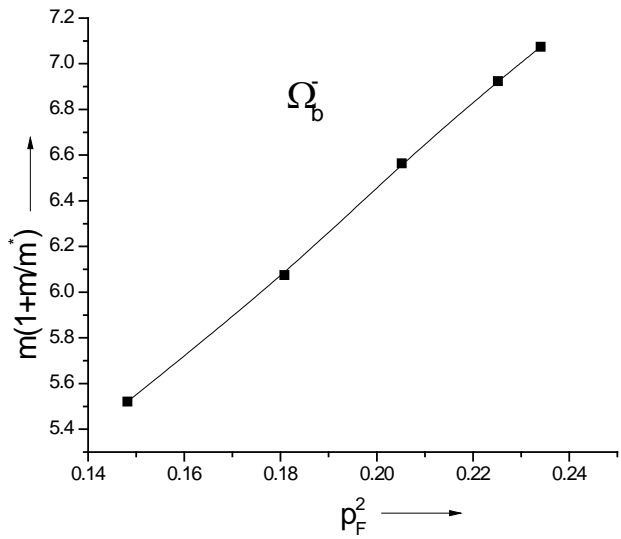

Figure 14. Variation of $p_F^2$ versus $m\left(1 + m/m^*\right)$